\documentclass[twocolumn,epsf]{jpsj3}

\title{Magnetic Field and Pressure Phase Diagrams of \\Uranium Heavy-Fermion Compound U$_2$Zn$_{17}$\thanks{J. Phys. Soc. Jpn. {\bf 80} 014706 (2011).}}

\author{Naoyuki \textsc{TATEIWA}$^{1}$\thanks{E-mail address: tateiwa.naoyuki@jaea.go.jp}, Shugo \textsc{IKEDA}$^{1,2}$, Yoshinori \textsc{HAGA}$^{1}$, Tatsuma D. \textsc{MATSUDA}$^{1}$, Etsuji \textsc{YAMAMOTO}$^{1}$, Kiyohiro \textsc{SUGIYAMA}$^{3}$, Masayuki \textsc{HAGIWARA}$^{4}$, Koichi \textsc{KINDO}$^{5}$, and Yoshichika \textsc{\=ONUKI}$^{1,3}$}
\inst{$^{1}$Advanced Science Research Center, Japan Atomic Energy Agency, Tokai, Ibaraki 319-1195\\
$^{2}$Graduate School of Material Science, University of Hyogo, Kamigori, Hyogo 678-1297\\ 
$^{3}$Graduate School of Science, Osaka University, Toyonaka, Osaka 560-0043\\
$^{4}$KYOKUGEN, Osaka University, Toyonaka, Osaka 560-8531\\
$^{5}$Institute for Solid State Physics, University of Tokyo, Kashiwa, Chiba 277-8581} 

\abst{We have performed magnetization measurements at high magnetic fields of up to 53 T on single crystals of a uranium heavy-fermion compound U$_2$Zn$_{17}$ grown by the Bridgman method. In the antiferromagnetic state below the N\'{e}el temperature $T_{\rm N}$ = 9.7 K,  a metamagnetic transition is found at $H_c$  $\simeq$ 32 T for the field along the [11$\bar{2}$0] direction ($a$-axis). The magnetic phase diagram for the field along the [11$\bar{2}$0] direction is given. The magnetization curve shows a nonlinear increase at $H_m$  $\simeq$ 35 T in the paramagnetic state above $T_{\rm N}$ up to a characteristic temperature $T_{{\chi}{\rm max}}$ where the magnetic susceptibility or electrical resistivity shows a maximum value. This metamagnetic behavior of the magnetization at $H_m$ is discussed in comparison with the metamagnetic magnetism of the heavy-fermion superconductors UPt$_3$, URu$_2$Si$_2$, and UPd$_2$Al$_3$. We have also carried out high-pressure resistivity measurement on U$_2$Zn$_{17}$ using a diamond anvil cell up to 8.7 GPa. Noble gas argon was used as a pressure-transmitting medium to ensure a good hydrostatic environment. The N\'{e}el temperature $T_{\rm N}$ is almost pressure-independent up to 4.7 GPa and starts to increase in the higher-pressure region. The pressure dependences of the coefficient of the $T^2$ term in the electrical resistivity $A$,  the antiferromagnetic gap $\Delta$, and the characteristic temperature $T_{{\rho}{\rm max}}$ are discussed. It is found that the effect of pressure on the electronic states in U$_2$Zn$_{17}$ is weak compared with those in the other heavy fermion compounds.
}

\kword{U$_2$Zn$_{17}$, magnetic phase diagram, pressure phase diagram}

\begin{document}
\maketitle

\section{Introduction}

Uranium intermetallic compounds exhibit unique electronic states such as magnetic orderings, heavy fermions,  and anisotropic superconductivity~\cite{onuki, flouquet1,lohneysen}. These properties are basically derived from the competition between the Ruderman-Kittel-Kasuya-Yosida (RKKY) interaction and the hybridization (Kondo) effect. The former interaction enhances the long-range magnetic order, where $5f$ electrons with magnetic moments are treated as localized electrons and the indirect $5f$-$5f$ interaction is mediated by the spin polarization of the conduction electrons. On the other hand, the latter effect quenches the magnetic moments of the localized $5f$ electrons by the spin polarization of the conduction electrons, leading to an extremely large density of states, called heavy fermions. 

 The application of pressure is a useful experimental method for controlling the magnetic RKKY interaction and hydridization effect. As pressure is applied to some compounds with magnetic orderings, the magnetic ordering temperature $T_{mag}$ decreases and becomes zero at a critical pressure $P_{c}$: ${T_{mag}}{\,}{\rightarrow}{\,}0$ at ${P}{\,}{\rightarrow}{\,}{P_{c}}$, where the pressure-induced superconductivity or non-Fermi liquid behavior appears. The pressure dependence of the magnetic ordering temperature $T_{mag}$ in cerium compounds is basically explained by the Doniach model, in which the magnetic ordering temperature $T_{mag}$ varies as a function of $|J_{cf}|D(\varepsilon_{\rm F})$\cite{doniach}, where $|J_{cf}|$ is a magnitude of the magnetic exchange interaction between the localized moment and the conduction electron spin, and $D(\varepsilon_{\rm F})$ is the electronic density of states at the Fermi energy $\varepsilon_{\rm F}$.  The pressure-induced superconductivity was discovered around $P_{c}$ in some cerium antiferromagnetic compounds such as CeCu$_2$Si$_2$, CeRh$_2$Si$_2$, CePd$_2$Si$_2$, CeIn$_3$, and CeRhIn$_5$~\cite{jaccard, movshovich, mathur,hegger}. On the other hand, pressure experiments on uranium compounds are small in number, and moreover, the pressure effect in uranium compounds seems to be small compared with that in cerium compounds of which the critical pressures $P_c$ are usually below 10 GPa. The pressure-induced superconductivity was only observed below $P_{\rm c}$ of the ferromagnetic state in UGe$_2$ and UIr~\cite{saxena,huxley,akazawa1,akazawa2}.

  In this study, we focus on the uranium heavy-fermion antiferromagnet U$_2$Zn$_{17}$ and studied its electrical and magnetic properties at a high magnetic field and a high pressure. U$_2$Zn$_{17}$ crystalizes in the rhombohedral Th$_2$Zn$_{17}$-type structure (space group $R$\={3}$m$)\cite{ott,fischer}.  At ambient pressure, U$_2$Zn$_{17}$ shows antiferromagnetic ordering at a N\'{e}el temperature $T_{\rm N}$ = 9.7 K with an ordered magnetic moment $\mu_{\rm ord}$ = 0.8 $\mu_{\rm B}$/U\cite{cox,aeppli}. The ordered moment is substantially below the paramagnetic moment of $3.15$ $\mu_{\rm B}$/U deduced from the high-temperature magnetic susceptibility measurement on a single crystal sample\cite{willis1}. The specific heat coefficient $C/T$ shows a large value of about 500 mJ/K${^2}{\cdot}$molU above $T_{\rm N}$ but is reduced to about $\gamma$  = 200 mJ/K${^2}{\cdot}$mol at $T$ ${\ll}$ $T_{\rm N}$~\cite{ott}. These results reveal the heavy-fermion nature of an itinerant-5$f$ electronic state in U$_2$Zn$_{17}$. A previous high-pressure experiment on U$_2$Zn$_{17}$ showed that the antiferromagnetic ordering temperature $T_{\rm N}$ increases slightly with increasing pressure from $T_{\rm N}$ = 9.70 K at 1 bar to 9.85 K at 1.72 GPa~\cite{thomspon1,thompson2}. In this study, we have measured the magnetization under high magnetic field, as well as performed the resistivity measurement under high pressures of up to 9 GPa using a diamond anvil cell.  
  
   \section{\label{sec:level1}Experimental Methods}
 A single-crystal sample of U$_2$Zn$_{17}$ was obtained by the Bridgman method with a W crucible  sealed with argon gas. The crucible was kept at 950-1050 $^{\circ}$C for 12 h and then cooled down slowly at a constant rate. The crystal structure was investigated by single-crystal X-ray diffraction techniques using an imaging plate (IP) area detector (Rigaku Corporation) with Mo $K{\alpha}$ radiation at room temperature.

 The electrical resistivity at both ambient and high pressures was measured by the four-probe DC method in the temperature range from 2 to 300 K. The magnetic susceptibility and magnetization were measured using a commercial superconducting quantum interference device (SQUID) magnetometer in the temperature range from 2 to 300 K.  The high-field magnetization was measured by the standard pick-up coil method at the High-Magnetic-Field Laboratory, KYOKUGEN, Osaka University, using a long-pulse magnet with a pulse duration of 20 ms.

      For the high-pressure study, a small sample was cut and polished to 180$\times$50$\times$20 $\mu$m$^3$. Four gold-wires 10 $\mu$m in diameter were bonded to the sample using silver paste. We used a diamond anvil cell of the Dunstan and Spain-type~\cite{dunstan1,dunstan2}. The sample and small ruby chips were placed in a sample hole 400 $\mu$m in diameter of a stainless-steel gasket in DAC. The culet-size of the diamonds is 800 $\mu$m. The electrodes are insulated from the metal gasket using a mixture of Al$_2$O$_3$ powder and stycast 1266~\cite{dunstan1,dunstan2,thomasson,knebel}. For a pressure-transmitting medium, we used the noble gas argon (Ar), which is known to provide a good hydrostatic condition up to 10 GPa at room temperature~\cite{bell,liu}. We confirmed that argon is also appropriate as the pressure-transmitting medium in the cryogenic experiment under high pressure~\cite{tateiwa}.  Argon can be liquefied at 87 K. It was loaded into the sample chamber of DAC using a purpose-built cryogenic device. The pressure was determined by the ruby fluorescence method at room temperature and 4.2 K~\cite{forman,barnett,piermarini1}. We used the hydrostatic ruby pressure scale obtained by Zha {\it et al}~\cite{zha}. Some ruby chips with a diameter less than 5 $\mu$m are placed in the sample chamber. Note that the pressure does not change significantly during the cooling process for the present DAC. The difference between the pressure at 4.2 and 300 K is less than 5$\%$. This is different from the other types of diamond anvil cells where the pressure at low temperatures is usually about 1 GPa higher or lower than that at room temperature. In the following, the pressure determined at 4.2 K will be shown.

 \begin{figure}[t]
\begin{center}
\includegraphics[width=8cm]{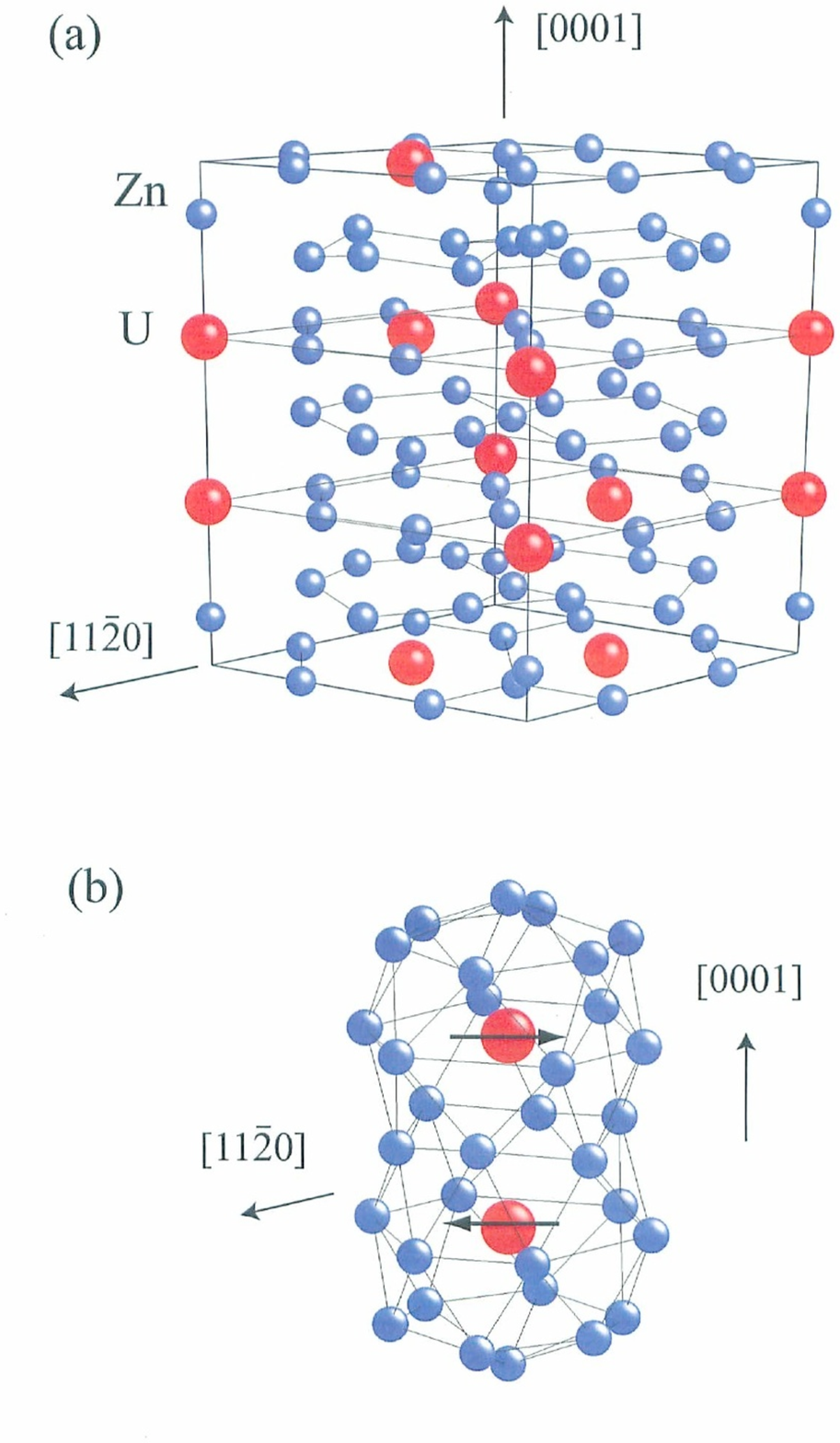}
 \end{center}
\caption{\label{fig:epsart} (Color online) Crystal structure in U$_2$Zn$_{17}$.}
\end{figure}

\section{Results and Discussion}
\subsection{\label{sec:level2}Crystal structure of U$_2$Zn$_{17}$}
 Figure 1(a) and 1(b) show the crystal structure of U$_2$Zn$_{17}$. Crystallographic parameters for U$_2$Zn$_{17}$ at room temperature is shown in Table I. The lattice parameters are  $a$ =8.9830(4){\AA} and $c$ =  13.1800(9) {\AA} at room temperature. The crystallographic parameters are consistent with the previous study within experimental error~\cite{siegrist1}. It is noted that the atomic coordinates in Table I are standardized using {\it STRUCTURE TIDY}~\cite{gelato}. The unit cell shown in Fig. 1 (a) containing 6 formula units (114 atoms) appears complicated.  To illustrate the local environment around the uranium site, we picked up neighboring Zn atoms, as shown in Fig. 1(b).  The uranium site has 19 Zn neighbors, forming a nearly spherical cage, with an open space toward the other uranium site connected to the next Zn cage.  The center of mass of the unit consisting of 2 U and 32 Zn atoms shown in Fig. 1(b) is located at (0,0,1/2) and equivalent positions.
The magnetic moments of the U ions in the cage, lying in the (0001) plane, are antiferromagnetically coupled in the ordered state below $T_{\rm N}$\cite{cox,aeppli}. The direction of the moment in the basal plane is not determined.

 \begin{table}
\caption{Crystallographic parameters for U$_2$Zn$_{17}$ at room temperature in the hexagonal setting (space group $R$\={3}$m$) with lattice parameters $a$ =8.9830(4){\AA} and $c$ =  13.1800(9) {\AA}. The conventional unweighted and weighted agreement factors of $R_1$ and $wR_2$ are 7.9 and 19$\%$, respectively. }
\label{t1}
\begin{center}
 \begin{tabular}{cccccc}
\hline
\multicolumn{1}{c}{Atom} & \multicolumn{1}{c}{Site} &\multicolumn{1}{c}{{\it x}} &\multicolumn{1}{c}{{\it y}} &\multicolumn{1}{c}{{\it z}} &\multicolumn{1}{c}{{\it B$_{\rm eq}$} ({\AA}$^2$)}\\
\hline
Zn(1) & 18h &  0.4956(3)  &  0.50438(16)   &  0.1526(2)   & 0.76(6) \\
Zn(2) &$~$18f  &  0.2965(3)  &  0   &  0   & 0.86(6)  \cr
Zn(3) &$~$9d  &  1/2 &  0   &  1/2   &0.70(7)  \cr
Zn(4) &  6c  &  0  &  0   &  0.0999(4)   &  0.80(8)  \cr
U  &  6c  &  0 &  0   &  0.33611(10)   &  0.51(5)  \cr
\hline
\end{tabular}
\end{center}
\end{table}

 \subsection{Resistivity and magnetic susceptibility at 1 bar}
    \begin{figure}[b]
\begin{center}
\includegraphics[width=8cm]{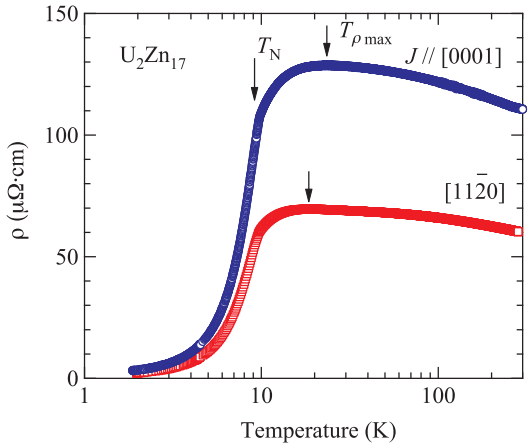}
 \end{center}
\caption{\label{fig:epsart}(Color online) Temperature dependences of the electrical resistivity at 1 bar for currents along $J$ $\|$ [11$\bar{2}$0] and [0001] directions in U$_2$Zn$_{17}$.}
\end{figure}     

  Figure 2 shows the logarithmic-scale of the temperature dependences of the electrical resistivity $\rho$ for the currents along the $J$ $\|$ [11$\bar{2}$0] ($a$-axis) and [0001] ($c$-axis) directions. The values of the residual resistivity ratio (RRR = ${\rho}_{\rm RT}/{\rho}_{\rm 0}$) are 75 for $J$ $\|$ [11$\bar{2}$0] and 81 for $J$ $\|$ [0001], where ${\rho}_{\rm RT}$ and ${\rho}_{\rm 0}$ are the resistivity at room temperature and the residual resistivity, respectively, indicating a comparably high quality of the present samples. The resistivity $\rho$ increases with decreasing temperature with a broad maximum at $T_{{\rho}max}$ = 18.7 and 23.5 K for $J$ $\|$ [11$\bar{2}$0] and [0001], respectively.  The resistivity shows a sharp kink at the N\'{e}el temperature $T_{\rm N}$ = 9.65 K and decreases steeply with decreasing temperature.  We define $T_{\rm N}$ as the peak position in the temperature dependence of  $d{^2}{\rho}/dT{^2}$.  The overall feature of the temperature dependence of the resistivity is roughly consistent with that reported in a previous study using a polycrystal sample~\cite{ott}. Thus far, the temperature dependence of the resistivity using a single crystal sample was reported only for $J$ $\|$ [0001]~\cite{siegrist2}. The resistivity $\rho$ for $J$ $\|$ [11$\bar{2}$0] is found to be approximately half as small as that for $J$ $\|$ [0001] above $T_{\rm N}$.

 \begin{figure}[t]
\begin{center}
\includegraphics[width=8cm]{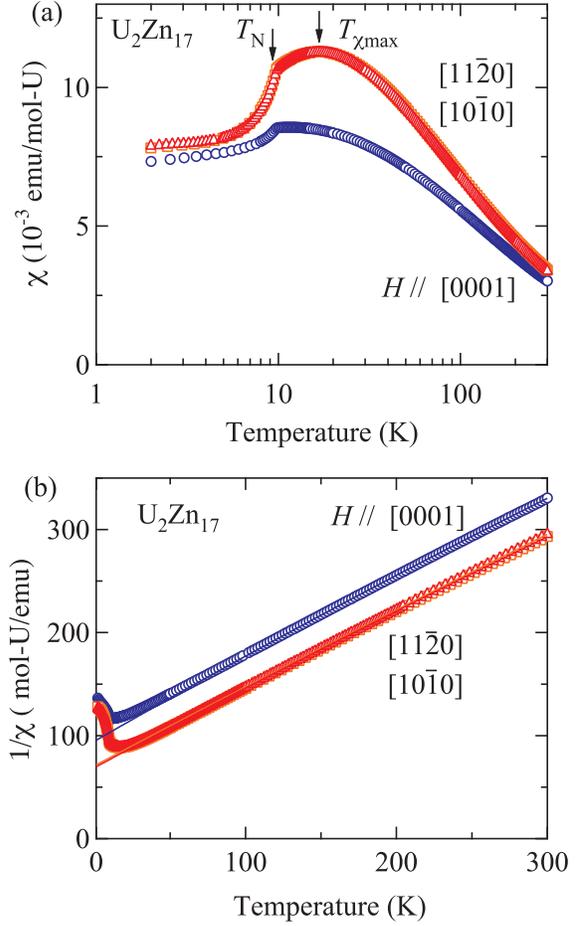}
 \end{center}
\caption{\label{fig:epsart}(Color online)Temperature dependences of (a) the magnetic susceptibility $\chi$ and (b) the inverse susceptibility 1/$\chi$ for the magnetic fields along [11$\bar{2}$0], [10$\bar{1}$0], and [0001] directions in U$_2$Zn$_{17}$.}
\end{figure} 
  
 Figure 3(a) shows the logarithmic-scale of the temperature dependences of the magnetic susceptibility ${\chi}$ for the magnetic fields along the [11$\bar{2}$0], [10$\bar{1}$0], and [0001] directions. The magnetic susceptibilities ${\chi}$ for $H$ $\|$ [11$\bar{2}$0] and [10$\bar{1}$0] have broad maxima at $T_{{\chi}{\rm max}}$ $\simeq$ 17 K, which is close to $T_{{\rho}{\rm max}}$ at which the resistivity shows a maximum. $T_{{\chi}{\rm max}}$ or $T_{{\rho}{\rm max}}$ corresponds to the characteristic temperature $T_{0}$ of the electronic state in U$_2$Zn$_{17}$. The susceptibility ${\chi}$ shows a sharp kink at the antiferromagnetic transition temperature $T_{\rm N}$ = 9.8 K and decreases steeply below $T_{\rm N}$. 
  
 The inverse magnetic susceptibility $1/{\chi}$ follows the Curie-Weiss law above 40 K for $H$ $\|$ [11$\bar{2}$0] and [10$\bar{1}$0], and above 100 K for  $H$ $\|$ [0001], as shown in Fig. 3(b). The effective paramagnetic moments $\mu_{\rm eff}$ and the Curie-Weiss temperatures $\Theta$ are = 3.19 ${\mu_{\rm B}}$/U  and - 91 K for $H$ $\|$  [11$\bar{2}$0],  3.15 $\mu_{\rm B}$/U and -87 K for $H$ $\|$ [10$\bar{1}$0] , and 2.99 $\mu_{\rm B}$/U and $\Theta$ = -109 K for $H$ $\|$ [0001], respectively. These values of $\mu_{\rm eff}$ for $H$ $\|$ [11$\bar{2}$0] and [10$\bar{1}$0] are roughly similar to the values for a free U ion value of 3.6 $\mu_{\rm B}$/U in the $5f^2$ and $5f^3$ configurations.

 The anisotropy of $\chi$ for three axes becomes smaller with decreasing temperature in the antiferromagnetic ordered state below $T_{\rm N}$. In particular, there is no significant difference between the temperature dependences of $\chi$ for $H$ $\|$ [11$\bar{2}$0] and [10$\bar{1}$0] perpendicular to the [0001] direction. A previous work has shown the magnetic susceptibilities for the fields parallel and perpendicular to the [0001] direction~\cite{willis1}. The result is basically consistent with the result of the present work. In the present study, it was clarified that there is no anisotropy of $\chi$ inside the (0001) plane even below $T_{\rm N}$. It is noted that the value of the critical exponent of the magnetic Bragg intensity $\beta$ (= 0.36 $\pm$ 0.02) in the neutron scattering experiment for U$_2$Zn$_{17}$ is between those expected for three-dimensional Heisenberg ($\beta$ = 0.367) and $XY$ ($\beta$ = 0.345) magnets~\cite{broholm,aeppli}. The anisotropy of the antiferromagnetic state is weak in U$_2$Zn$_{17}$.

  \begin{figure}[t]
\begin{center}
\includegraphics[width=8cm]{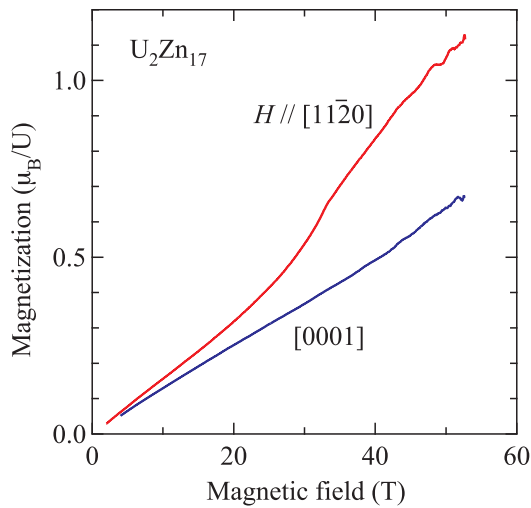}
 \end{center}
\caption{\label{fig:epsart}(Color online) Magnetization curves in the magnetic field along [11$\bar{2}$0] and [10$\bar{1}$0] directions at 1.3 K in U$_2$Zn$_{17}$.}
\end{figure} 

   \begin{figure}
\begin{center}
\includegraphics[width=8cm]{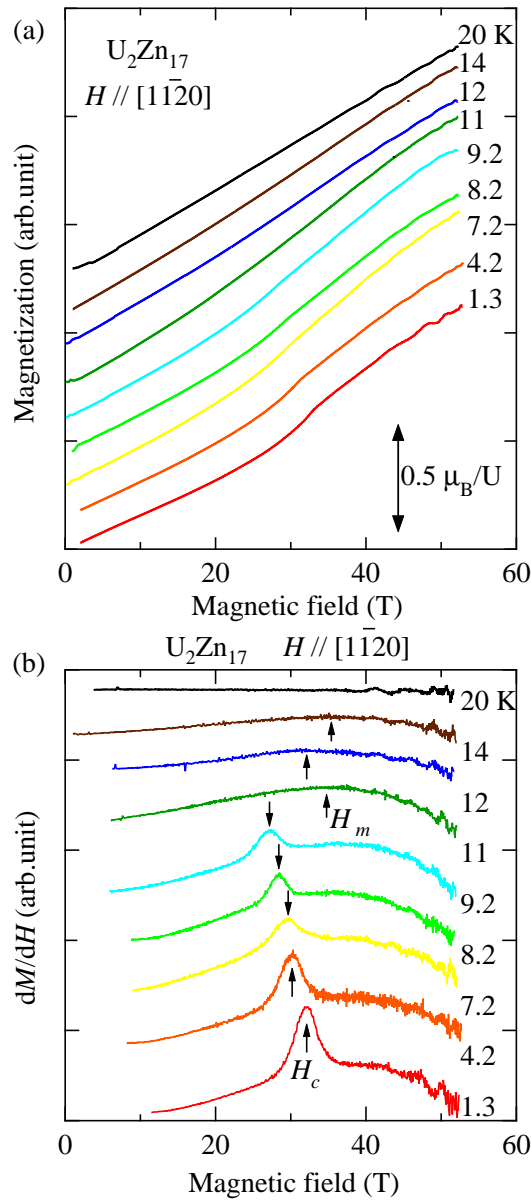}
 \end{center}
\caption{\label{fig:epsart}(Color online) (a) Magnetization curves at various temperatures and (b) differential magnetization curves for the field along the [11$\bar{2}$0]  direction in U$_2$Zn$_{17}$.}
\end{figure}  

  \subsection{High-magnetic-field experiment}
      \begin{figure}[t]
\begin{center}
\includegraphics[width=7cm]{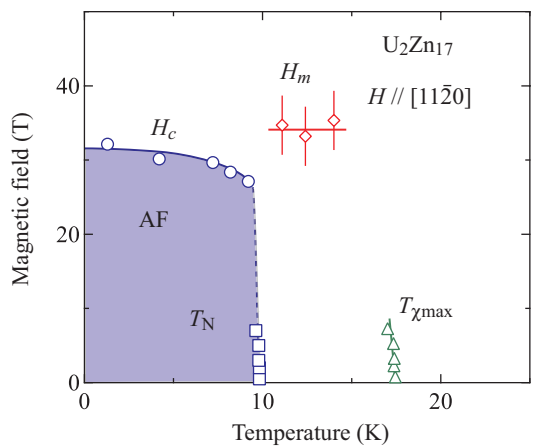}
 \end{center}
\caption{\label{fig:epsart}(Color online) Magnetic phase diagram of U$_2$Zn$_{17}$ for the field along the [11$\bar{2}$0] direction. Solid and dotted lines are a guide to the eyes.}
\end{figure}  

  Figure 4 shows the magnetization curves for the field along the [11$\bar{2}$0] and [10$\bar{1}$0] directions at 1.3 K. There is no strong anisotropy in the magnetization processes at a low magnetic field. The magnetization for $H$ $\|$ [11$\bar{2}$0] increases approximately linearly with increasing magnetic field and shows a metamagnetic transition at $H_{\rm c}$ = 33 T. With further increasing field, the magnetization increases monotonically and amounts to 1.1 ${\mu}_{\rm B}$/U at 50 T. The present value is larger than the antiferromagnetic ordered moment (0.8 ${\mu}_{\rm B}$), indicating the itinerant band magnetism of $5f$ electrons in U$_2$Zn$_{17}$.  The magnetization for $H$ $\|$ [0001] increases linearly as a function of magnetic field and starts to deviate upward from about 42 T. This result suggests that the metamagnetic transition also exists for $H$ $\|$ [0001] at a magnetic field higher than 52 T, the highest magnetic field in the present study.

 Figure 5 (a) shows the magnetization curves in the magnetic field along the [11$\bar{2}$0] direction at various temperatures. The corresponding field derivatives of the magnetization curve, d$M$/d$H$, are shown in Fig. 5 (b). The metamagnetic transition at $H_{c}$ becomes broad with increasing temperature up to 9.2 K, just below $T_{\rm N}$ = 9.65 K.  
  At 11, 12, and 14 K, the slope of the magnetization curves shows another metamagnetic behavior at $H_{m}$. In fact, there appear broad peaks at $H_{m}$ in the d$M$/d$H$ curves at these temperatures, as shown in Fig. 5 (b). At 20 K, the magnetization increases linearly. 
       
  Figure 6 shows the magnetic phase diagram for $H$ $\|$ [11$\bar{2}$0]. The data obtained from the high-field magnetization measurement are shown by open circles and diamonds. The field dependences of $T_{\rm N}$ and $T_{{\chi}{\rm max}}$, determined by the SQUID magnetization measurement, are shown by open squares and triangles, respectively. The metamagnetic transition field $H_{c}$ in the antiferromagnetic order state decreases from 32.5 T at 1.3 K to 27.8 T at 9.2 K. A curve connecting the $H_{c}$ data seems to reach the phase boundary of $T_{\rm N}$ below 7 T. It seems that the metamagnetic behavior at $H_{m}$ in the paramagnetic state appears in the temperature region between $T_{\rm N}$ and $T_{{\chi}{\rm max}}$.   
  
       \begin{figure}[t]
\begin{center}
\includegraphics[width=8cm]{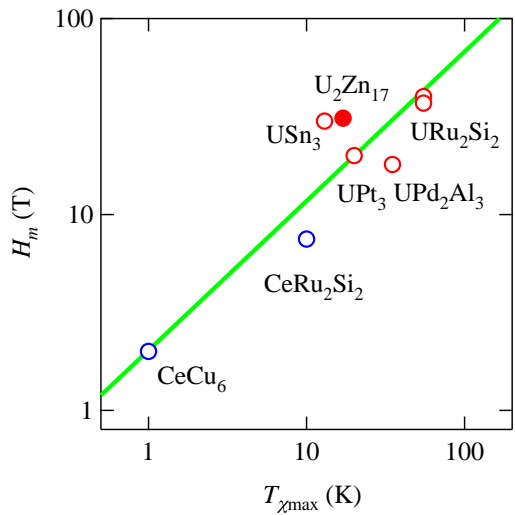}
 \end{center}
\caption{\label{fig:epsart}(Color online) Metamagnetic anomaly fields $H_{m}$ of several heavy-fermion compounds, shown in logarithmic scale. A closed circle corresponds to that in the case of U$_2$Zn$_{17}$.}
\end{figure}  
  
  We suggest that the metamagnetic behavior at $H_{m}$ in the paramagnetic state of U$_2$Zn$_{17}$ is similar to those observed for heavy-fermion compounds such as UPt$_3$, URu$_2$Si$_2$, and UPd$_2$Al$_3$ below $T_{{\chi}{\rm max}}$, where the magnetic susceptibility shows a maximum~\cite{sugiyama1,sugiyama2,sugiyama3}. It is basically supposed that the behavior is associated with the change of the hybridization effect between conduction electrons with a wide energy band and almost localized $f$-electrons~\cite{onuki}. Almost localized $f$-electrons become itinerant with decreasing temperature through the many-body effect. The crossover from localized to itinerant occurs at a characteristic temperature $T_0$, corresponding to $T_{{\chi}{\rm max}}$ or $T_{{\rho}{\rm max}}$, where the magnetic susceptibility or the electrical resistivity has a maximum. The metamagnetic behavior appears at $H_{\rm m}$, where the relation of $k_{\rm B}T_{{\chi}{\rm max}}  \simeq {\mu_{\rm B}}H_{\rm c}$ is realized by applying magnetic field, as shown in Fig. 7. The relation of $T_{{\chi}{\rm max}}$ = 17 K and $H_m$ = 32 T in U$_2$Zn$_{17}$ is shown in the Fig. 7 by a closed circle, approximately consistent with this relation.
  
   Various microscopic theoretical studies have been performed on the metamagnetic behavior of the magnetization in heavy-fermion compounds. Miyake and Kuramoto have calculated the magnetization process using a semi-phenomenological model called the duality model of heavy fermions on the periodic Anderson lattice model~\cite{kuramoto,miyake}. In the model, the metamagnetic behavior takes place when the second derivative of the density of states is positive and the coupling between itinerant and localized parts of $f$ electrons is large. From a different point of view, it was proposed that the anisotropy of the hybridization matrix element yields the characteristic shape of the density of states that plays a major role in the metamagnetism~\cite{ono,ohara}. Ohkawa and coworkers clarified the spin-lattice effect cooperating with the ferromagnetic exchange interaction causes the metamagnetic behaviors~\cite{ohkawa1,ohkawa2,ohkawa3}. Although a final consensus has not been established yet, the metamagnetic behavior in the heavy-fermion system is one of the important issues in $f$-electron magnetism.  We hope that the present observation in U$_2$Zn$_{17}$ stimulates future studies on the issue.

  \subsection{High-pressure experiment}
    \begin{figure}[t]
\begin{center}
\includegraphics[width=8cm]{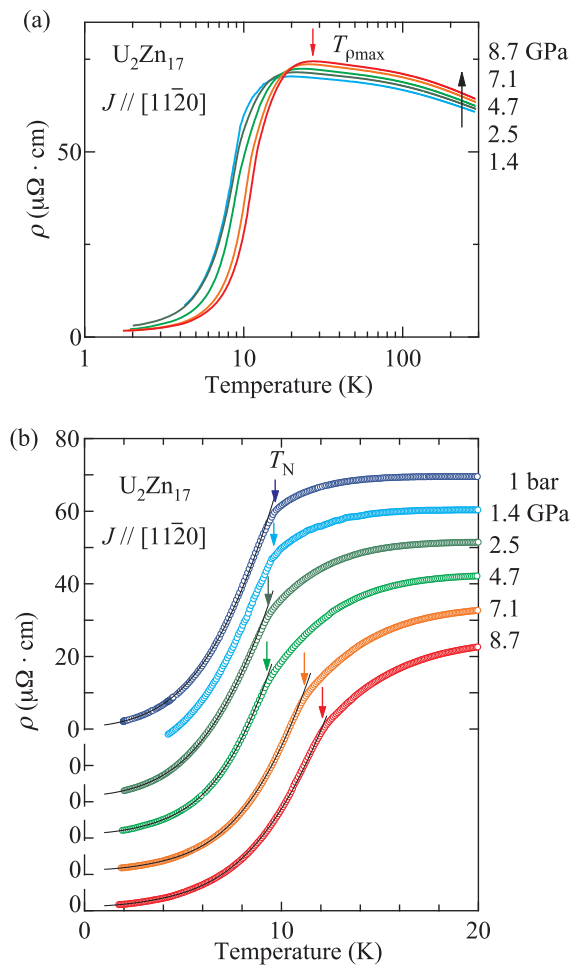}
 \end{center}
\caption{\label{fig:epsart}(Color online) (a)Temperature dependences of the electrical resistivity $\rho$ and (b) low-temperature resistivity at several pressures in U$_2$Zn$_{17}$.}
\end{figure}    
   Figure 8 (a) shows the logarithmic temperature dependence of the electrical resistivity $\rho$ in the current parallel to the [11$\bar{2}$0] direction under high pressures.  The resistivity at room temperature increases slightly with increasing pressure. The characteristic temperature $T_{{\rho}{\rm max}}$, where the resistivity shows a maximum, is shifted to the higher temperature side with increasing pressure. $T_{{\rho}{\rm max}}$ is 27.6 K at 8.7 GPa.

 To clarify the behavior of the resistivity ${\rho}$ around the magnetic ordering temperature $T_{\rm N}$, we show the low-temperature resistivity in Fig. 8 (b), where the N\'{e}el temperature $T_{\rm N}$ is shown by an arrow. The pressure dependence of $T_{\rm N}$ is shown in Fig. 9. The N\'{e}el temperature $T_{\rm N}$ is almost pressure-independent up to 4.7 GPa. Above the pressure, $T_{\rm N}$ starts to increase with increasing pressure. $T_{\rm N}$ is 12.2 K at 8.7 GPa. A characteristic feature is that the sign of $dT_{\rm N}/dT$ changes at approximately 5 GPa. The present result is roughly consistent with that of the previous study up to 1.7 GPa~\cite{thomspon1,thompson2}. 

\begin{figure}[t]
\begin{center}
\includegraphics[width=8cm]{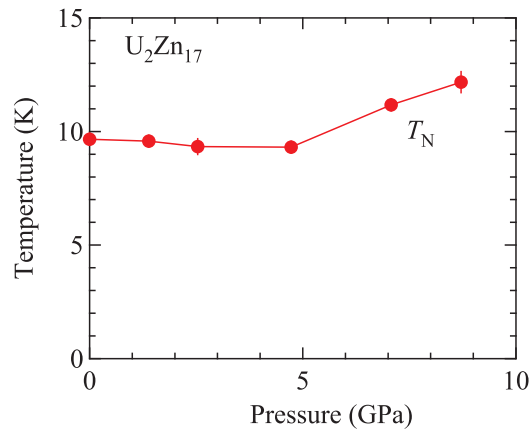}
 \end{center}
\caption{\label{fig:epsart}(Color online) Pressure dependence of the N\'{e}el temperature $T_{\rm N}$ in U$_2$Zn$_{17}$.}
\end{figure}   
  The resistivity below $T_{\rm N}$ is analyzed using the antiferromagnetic magnon model described as
  \begin{equation}
\rho = {\rho}_0 + AT^2+BT(1+{\frac{2T}{\Delta}}){\rm exp}(-{\frac{\Delta}{T}}),
\end{equation}
where the third term corresponds to the contribution of the electron scattering by an antiferromagnetic magnon with an energy gap ${\Delta}$, which was used in the analyses of URu$_2$Si$_2$ and CePd$_2$Si$_2$~\cite{palstra2,raymond}.  A fit of the resistivity data is shown by solid lines in Fig. 8 (b). The pressure dependences of the obtained parameters $A$ and ${\Delta}$ are shown in Fig. 10 (a). $A$ decreases simply with increasing pressure, from 0.35 ${\mu}{\Omega}{\cdot}$cm/K$^2$ at 1 bar to 0.14 ${\mu}{\Omega}{\cdot}$cm/K$^2$. The corresponding ${\gamma}$ values are estimated as 190 and 120 mJ/K${^2}{\cdot}$mol at 1 bar and 8.7 GPa, respectively, using the Kadowaki-Woods relation ($A/{{\gamma}^2} = 1.0 {\times} 10^{-5}$)~\cite{kadowaki}. The estimated ${\gamma}$ at 1 bar is consistent with the observed value  of about $\simeq$ 200 mJ/K${^2}{\cdot}$molU. It is suggested that ${\gamma}$ decreases with increasing pressure. ${\Delta}$ increases monotonically from 19 K at 1 bar to 33 K at 8.7 GPa, indicating that the antiferromagnetic state is enhanced with increasing pressure.

 \begin{figure}[t]
\begin{center}
\includegraphics[width=8cm]{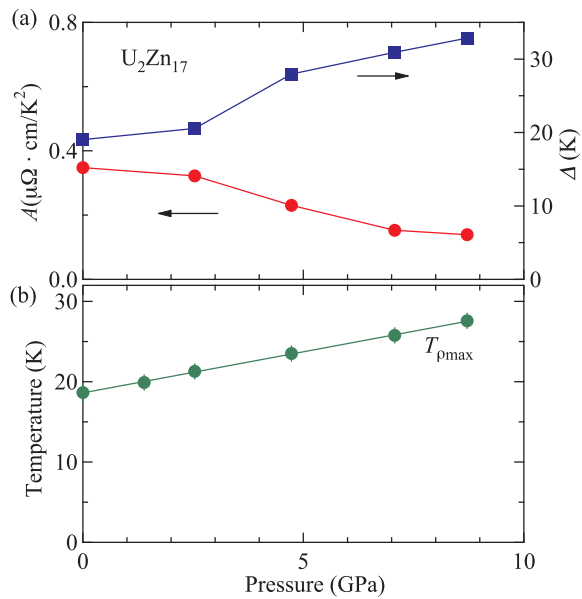}
 \end{center}
\caption{\label{fig:epsart}(Color online) Pressure dependences of (a) the coefficient of $T^2$ term of the resistivity $A$(left side) and the antiferromagnetic gap $\Delta$ (right side), and (b) the characteristic temperature $T_{{\rho}{\rm max}}$ where the resistivity $\rho$ shows a maximum value in U$_2$Zn$_{17}$.}
\end{figure} 

Very recently, Sidorov, {\it et al.} performed the resistivity and ac heat capacity measurements on U$_2$Zn$_{17}$ up to 5.5 GPa~\cite{sidorov}. The reported pressure dependence of $T_{\rm N}$ is roughly consistent with that observed in the present study. In the study, two successive magnetic transitions were observed in the pressure range of 2.64 - 3.25 GPa. It was concluded that the antiferromagnetic ground state changes to a new antiferromagnetic phase at approximately 2.4 GPa. Since we have not investigated the pressure dependence of $T_{\rm N}$ in detail at approximately this pressure, the change of the magnetic ground state is not discussed within the present data. We only mention the possibility that the slight change in the pressure dependences of $A$ and ${\Delta}$ at approximately 3 GPa shown in Fig. 10 (a) is due to the appearance of the pressure-induced new magnetic phase revealed by our work. 

  Figure 10 (b) shows the pressure dependence of $T_{{\rho}{\rm max}}$. This characteristic temperature $T_{{\rho}{\rm max}}$ varies linearly as a function of pressure, shown as a solid straight line in the Fig. 10 (b). The pressure derivative of ${\partial T_{{\rho}{\rm max}}}/{\partial P}$ is 1.0 K/GPa. Noted that $T_{{\rho}{\rm max}}$ corresponds to the characteristic temperature $T_0$ of the electronic state in U$_2$Zn$_{17}$. The Gr{\"u}neisen parameter ${\it {\Gamma}}_{T_{\rm 0}}$ for $T_{\rm 0}$ is written as follows: 
  \begin{equation}
  {\it {\Gamma}}_{T_{\rm 0}} = -{\frac{\partial\,{\rm ln}T_{\rm 0}}{\partial\,{\rm ln}V}} = B\frac{1}{T_{\rm 0}}{\frac{\partial\,T_{\rm 0}}{\partial\,P}},
  \end{equation}
where $B$ is the bulk modulus, estimated to be 83 GPa at 297 K for U$_2$Zn$_{17}$ by ultrasound measurement~\cite{migliori}. The Gr{\"u}neisen parameter ${\it {\Gamma}}_{T_{\rm 0}}$ is 4.6 in U$_2$Zn$_{17}$. This value is small compared with those of heavy-fermion compounds, where the Gr{\"u}neisen parameter is usually about one or two orders of magnitude larger than those of ordinary metals~\cite{thompson2,kagayama1,kagayama2}. For example, the Gr{\"u}neisen parameters for the characteristic temperature $T_0$ of the heavy-fermion superconductors CeCu$_2$Si$_2$, UBe$_{13}$, UPt$_3$, and URu$_2$Si$_2$ are 22, 103, 76, and 19, respectively~\cite{thompson2,kagayama1}. In Fig. 11, we plot the relations of the Gr{\"u}neisen parameter ${\it {\Gamma}}_{T_{\rm 0}}$ and the electronic specific heat coefficient ${\gamma}$ for several cerium and uranium heavy-fermion compounds~\cite{thompson2,kagayama1,kagayama2,link}.  
Theoretically, the ${\gamma}$ value correlates with the Gr{\"u}neisen parameter enhanced by the many-body effect in the Kondo lattice system~\cite{hong,flouquet2}. It is supposed that the Gr{\"u}neisen parameter of U$_2$Zn$_{17}$ is unusually small when the large $\gamma$ value is taken into account. The Gr{\"u}neisen parameter ${\it {\Gamma}}_{A}$ for $A$ in the resistivity is estimated as 2.5 in the low-pressure region. This value is also small compared with those of UBe$_{13}$ and UPt$_3$, where the values of ${\it {\Gamma}}_{A}$ are 17 and 61, respectively~\cite{thompson2}. These small values of the Gr{\"u}neisen parameters suggest that the electronic state in U$_2$Zn$_{17}$ is not so sensitive to the change of the lattice parameters compared with those of the other heavy-fermion compounds with the order of $\gamma$ = 100 mJ/K${^2}{\cdot}$mol.

      \begin{figure}[t]
\begin{center}
\includegraphics[width=8cm]{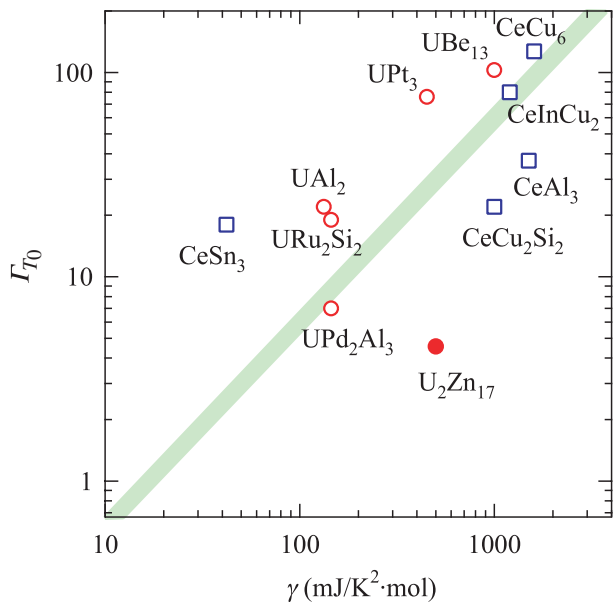}
 \end{center}
\caption{\label{fig:epsart}(Color online) Gr{\"u}neisen parameter ${\it {\Gamma}}_{T_{\rm 0}}$ vs linear heat capacity coefficient $\gamma$ for several cerium and uranium compounds~\cite{thompson2,kagayama1,kagayama2}. The bold line is a guide to the eye.}
\end{figure} 

  It is interesting to note that the antiferromagnetic state in U$_2$Zn$_{17}$ is very sensitive to a small amount of substitution on the Zn site~\cite{willis2}. The antiferromagnetic transition temperature $T_{\rm N}$ is strongly depressed to below 1.5 K with the substitution of only 2\% Cu on the Zn site.  On the other hand, the magnetic ordered state does not seem to be easily destroyed by high pressure. A much higher pressure far above 10 GPa seems to be needed to suppress the magnetic ordered state. 
 
  We compare the pressure effect on the antiferromagnetic state in U$_2$Zn$_{17}$ with the cerium antiferromagnetic compounds CeIn$_3$ and CeRhIn$_5$ whose bulk moduli of 67 and 78, respectively, are close to that of U$_2$Zn$_{17}$~\cite{vedel,kumar}. Both CeIn$_3$ and CeRhIn$_5$ order antiferromagnetically at $T_{\rm N}$ =10 K and 3.8 K, respectively, at ambient pressure. Under high pressure, the antiferromagnetic ordering temperature $T_{\rm N}$ becomes 0 K at critical pressures $P_{c}$ of 2.1 and 2.5 GPa, respectively, and the ground state changes into the superconducting one at approximately $P_c$. The magnetic to non-magnetic transition takes place below 3 GPa in both compounds whose compressibility $\beta$ (= $1/B$) is similar to that of U$_2$Zn$_{17}$. On the other hand, the antiferromagnetic ordered state in U$_2$Zn$_{17}$ is not easily destroyed, but is stabilized under high pressures of up to 9 GPa. The pressure effect on the antiferromagnetic phase as well as on the paramagnetic electronic state in U$_2$Zn$_{17}$ is thus highly different from those of the heavy-fermion cerium compounds described basically by the Doniach model for the Kondo lattice. The weak pressure effect on $T_{\rm N}$ may be ascribed to the itinerant character of the 5$f$ electrons in U$_2$Zn$_{17}$, as discussed in uranium monochalcogenides UX(X=Te, Se, S)~\cite{link1,link2,huang,cornelius}. The pressure dependence of the Curie temperature $T_{\rm C}$ in uranium monochalcogenides deviates from that in the Doniach model, but is explained by the spin fluctuation theory of an itinerant 5$f$ electron system based on the Hubbard model~\cite{cooper1,cooper2}.  In the theory, the pressure-induced increase in the hydridization between 5$f$ and conduction electrons strengthens the exchange interaction $J$ between uranium ions, but it also decreases the 5$f$ spectral weight (magnetic moment) at the uranium site. The model predicts a much more gradual demagnetization of the itinerant magnetic system under higher pressure than the Doniach picture. The present weak pressure dependence of  $T_{\rm N}$ in U$_2$Zn$_{17}$ may reflect the itinerant character of the 5$f$ electrons in the compound, which is different from the 4$f$ electron system in cerium compounds. It is interesting to note high-pressure studies of the well-known heavy-fermion antiferromagnet UCu$_5$ where the antiferromagnetic phase transition takes place at $T_{\rm N}$ = 15.9 K.  It was revealed that the N\'{e}el temperature $T_{\rm N}$ of UCu$_5$ increases with increasing pressure very gradually at a rate of 0.33 K/GPa and that the magnetic phase exists even at 13 GPa~\cite{thompson3,nakashima}. It is suggested that the pressure response of uranium heavy fermion compounds with the antiferromagnetic ground state generally differs from that of the 4$f$ electron system in cerium compounds.  
 
\section{Conclusions}

 We have performed a high-field magnetization experiment as well as a high-pressure experiment on single crystals of U$_2$Zn$_{17}$ grown by the Bridgman method. We also measured the electrical resistivity and magnetic susceptibility at ambient pressure. The experimental results are summarized as follows: \\
 1) Both the magnetic susceptibility $\chi$ and electrical resistivity $\rho$ at ambient pressure show a broad maximum at approximately 18 K in the paramagnetic state above $T_{\rm N}$, similarly to those observed in the heavy-fermion superconductors UPt$_3$, UPd$_2$Al$_3$, and URu$_2$Si$_2$, which is consistent with the results of the previous studies. The magnetic susceptibility is anisotropic between $H$ $\|$ [0001] and $H$ $\bot$ [0001], but the anisotropy of the susceptibility is not present betweeen $H$ $\|$ [11$\bar{2}$0] and $H$ $\|$ [10$\bar{1}$0]. The resistivity $\rho$ for $J$ $\|$ [11$\bar{2}$0] is approximately half as small as that for $J$ $\|$ [0001] above $T_{\rm N}$.\\
 2) In the antiferromagnetic state below $T_{\rm N}$, the metamagnetic transition is observed at $H_{c}$ = 30 T in the field along the antiferromagnetic easy axis of [11$\bar{2}$0]. The magnetic phase diagram for the field along the [11$\bar{2}$0] direction is given. The magnetization shows the metamagnetic behavior at $H_m$  $\simeq$ 35 T in the paramagnetic state above $T_{\rm N}$. We suggest that the behavior of the magnetization is the same as those observed in heavy-fermion compounds such as UPt$_3$, UPd$_2$Al$_3$, and URu$_2$Si$_2$. \\
3) From the high-pressure experiment using the diamond anvil cell, it was clarified that the antiferromagnetic ordering temperature $T_{\rm N}$ is almost pressure-independent up to 4.7 GPa and starts to increase in the higher-pressure region. The critical pressure for the magnetically ordered state seems to be far above 10 GPa.\\
4) The Gr{\"u}neisen parameter ${\it {\Gamma}}_{T_{\rm 0}}$ for the characteristic temperature $T_0$ was estimated to be 4.6 for U$_2$Zn$_{17}$ from the pressure dependence of $T_{{\rho}{max}}$, where the resistivity $\rho$ shows a maximum value. The value of the parameter is small compared with those of the other heavy fermion compounds, of which the electronic specific heat linear coefficient $\gamma$ is in the order of 100 mJ/K${^2}{\cdot}$mol, indicating a small response of the electronic state to a change of the lattice parameter. 

\section{Acknowledgements}

This work was financially supported by a Grant-in-Aid for Scientific Research on Innovative Areas ``Heavy Electrons (No. 20102002), Scientific Research S (No. 20224015), C (No. 21540373, 22234567), Specially Promoted Research (No. 20001004) and Osaka University Global COE Program (G10) from the Ministry of Education, Culture, Sports, Science and Technology (MEXT) and Japan Society of the Promotion of Science (JSPS).

\end{document}